\documentclass[aip, preprint]{revtex4}
\usepackage{graphicx,epsf,amssymb,amsmath,latexsym,epsfig}

\newcommand{\e}[1]{\mbox{e}^{#1}}

\newcommand{\eexp}{\mbox{e}^}
\newcommand\be{\begin{equation}}
\newcommand\ee{\end{equation}}
\newcommand\lab[1]{\label{eq:#1}}

\begin{document}

\title{Axions Scattering From a Quadrupole Magnetic Field}

\vspace{.3in}

\author{Eduardo I. Guendelman}
\email{guendel@bgu.ac.il}
\author{Shay Leizerovitch}
\email{laizerov@bgu.ac.il}
\author{Idan Shilon}
\altaffiliation{Present address: \textit{European Organization for Nuclear Research (CERN), CH-1211 Gen\`eve 23, Switzerland}}
\email{idan.shilon@cern.ch}
\affiliation{Physics Department, Ben Gurion University of the Negev, Beer-Sheva 84105, Israel}

\vskip.3in

\begin{abstract}

We study the 2D scattering of axions from an accelerator like quadrupole magnet using the eikonal approximation in order to learn whether or not such a setup could serve as a new possible method for detecting axions on terrestrial experiments. The eikonal approximation in 2D is introduced and explained. We also apply the eikonal approximation to two known cases in order to compare it with previous results, obtained using Born's approximation, and discuss its correctness. 
\end{abstract}

\maketitle

\setcounter{equation}{0}

\section{introduction}

The question of the scattering of axions from a quadrupole magnetic field has recently been proposed as a new possible method for detecting axions on terrestrial experiments such as the CAST experiment at CERN. Current axion detection experiments use a constant magnetic field, that can be generated by a particle accelerator dipole magnet for example, to trace QCD axions. In the case of CAST, these axions are assumed to be emerging from the sun with a mean energy which is estimated to be $E = 4.2\times 10^{3}$ eV \cite{ene}. These experiments use a standard 1D analysis (i.e the inverse Primakoff effect) \cite{sik} to evaluate the axion-photon reconversion. Therefore, it is an interesting question wether a magnetic field which is varying over the scattering region, and thus should be analyzed by a 2D formalism, could improve the probability of QCD axion detection in the near future. In this work we estimate the cross-section and conversion probability of solar axions that scatter along a long accelerator like quadrupole magnet, using a novel 2D scattering method which accounts for an axion-photon splitting and presented in the means of an axion-photon duality symmetry.

In a recent publication, two of us have studied the 2D scattering of axions from a magnetic field with cylindrical symmetry \cite{2d}. In this work, we expand our study to analyze the scattering process when the external magnetic field is generated by a magnetic quadrupole using the particle anti-particle 2D scattering formalism, introduced in \cite{duality} and reviewed in \cite{2d}. This new 2D formalism uses a duality symmetry between the axion field and the scattered component of the photon to define an axion-photon complex field as

\begin{equation}
\label{ }
  \Psi = \frac{(\phi + iA)}{\sqrt{2}} ~,
\end{equation}
 
 \noindent where $\phi$ is the axion field and $A$ is the $z$-polarization of the photon.  We focus here, as in \cite{2d}, on the case where an electromagnetic field with propagation along the $x$ and $y$ directions and an external magnetic field pointing in the $z$-direction are present. The magnetic field may have an arbitrary space dependence in $x$ and $y$, but it is assumed to be time independent. Hence, for the electric field component in the perpendicular direction to the plane we have $E_z = -\partial_{t}A$. In the case where the scattering process takes place in a medium, the photons acquire an effective mass. By matching the photon effective mass to the axion mass, the duality symmetry is again discovered \cite{2d}. However, for convenience let us neglect the axion mass and consider the experiment to be conducted in vacuum so we can write the lagrangian in terms of the new canonical variables $\Psi$ and its charge conjugate $\Psi^{*}$ 

\begin{equation}
\label{ }
	\mathcal{L} = \partial_{\mu}\Psi^{*}\partial^{\mu}\Psi -\frac{i}{2}\beta(\Psi^{*}\partial_{t}\Psi - \Psi\partial_{t}			\Psi^{*})~,
\end{equation}

\noindent where $\beta(x,y) = gB(x,y)$ with $B(x,y)$ being the external magnetic field and $\Psi^{*}$ is the charge conjugation of $\Psi$. From this we obtain the equation of motion for $\Psi$

\begin{equation}
\label{equation7}
	\partial_{\mu}\partial^{\mu}\Psi + i\beta\partial_{t}\Psi = 0 ~.
\end{equation}

\noindent We therefore have the magnetic field, or $\beta/2$ (the $U(1)$ charge), coupled to a charge density. Introducing the charge conjugation (C.C) \cite{part-antipart} (i.e $\Psi \rightarrow \Psi^{*}$) shows that the free part of the action is indeed invariant under C.C. When acting on the free vacuum the $A$ and $\phi$ fields give rise to a photon and an axion respectively, but in terms of the particles and antiparticles (defined in terms of  $\Psi$), we see that a photon is an antisymmetric combination of particle and antiparticle and an axion a symmetric combination, since

\be
\phi =\frac{1}{\sqrt{2}}(\Psi^{*} +\Psi) ~~\mbox{and} ~~A= \frac{1}{i\sqrt{2}}(\Psi - \Psi^{*})~.
\lab{part, antipart}
\ee

\noindent Hence, the axion is even under charge conjugation, while the photon is odd.
These two eigenstates of charge conjugation will propagate without mixing as long as no external magnetic field in the perpendicular direction to the eigenstates (i.e axion and photon) spatial dependence is applied.
The interaction with the external magnetic field is not invariant under C.C. In fact,
under C.C we can see that $S_I \rightarrow - S_I$, where $S_{I} = \int\mathcal{L}_{I}dxdydt$. Therefore, these symmetric and antisymmetric combinations, corresponding to axion and photon, will not be preserved in the presence of $B$  in the analog QED language, since the "analog external electric potential" breaks the symmetry between particle and antiparticle and therefore will not keep in time the symmetric or antisymmetric combinations. Moreover, if the corresponding external electric potential is taken to be a repulsive potential for particles, it will be an attractive potential for antiparticles, so the symmetry breaking is maximal.


\section{The Scattering Amplitude in a 2D Eikonal Approximation}

To apply the results of the previous section to some specific systems with magnetic field, we write separately the time and space dependence of the axion-photon field as $\Psi(\vec{r},t) = \mbox{e}^{-i\omega t}\psi_{k}(\vec{r})$, which yields
\begin{equation}
\label{equation of motion}
	(-\omega^{2} -\nabla^{2} +\omega\beta)\psi_{k} = 0 ~.
\end{equation}
\noindent In order to develop the space dependent term of the axion-photon field, let us consider a high energy, non-relativistic scattering. We assume that the wavelength of the $\psi$ field is short, i.e $kR\gg1$, where $R$ is the length scale of the scattering region and $k$ is the momentum of the incoming beam, and that $|V_0|/E\ll 1$, where $|V_0| = gB_{0}$ is the averaged magnitude of the potential over the scattering region and $E$ is the energy of the incoming beam. Under these assumptions we can address the problem by assuming small scattering angles and representing our equation in the integral from of the Lippman-Schwinger equation

\begin{equation}
\label{psi}
 \psi_k(r)= \e{i\vec{k}\cdot\vec{r}}+\int d^2r^\prime G_0^{(+)}(r,r^\prime)U(r^\prime)\psi_k(r^\prime)~,
\end{equation}

\noindent where $G_{0}^{(+)}(r,r^\prime)$ is Green's function given by

\be
\label{green function}
 G_0^{(+)} =\frac{1}{2\sqrt{2 \pi k|r-r'|}}\e{i(k|r-r^\prime|+\pi/4)} = \frac{1}{(2 \pi)^2}\lim_{\epsilon\to 0}\int d^2 k^\prime\frac{\e{ik^\prime (r-r^\prime)}}{k^{\prime 2}-k^2-i\epsilon}~,
\ee

\noindent where $k = \omega$ for a massless field. writing the spatial part ot the wave function $\psi_{k}(r)$ as

\begin{equation}
\label{psi2}
	 \psi_k(r)= \e{i\vec{k}\cdot\vec{r}}\phi(r) ~,
\end{equation}

\noindent and substituting into Eq.(\ref{psi}) yields an equation for $\phi(r)$ \cite{eik}

\begin{equation}
\label{phi}
\begin{array}{c}
	 \phi(r) = 1+\e{-i\vec{k}\cdot\vec{r}}\int d^2r^\prime G_0^{(+)}(r,r^\prime)U(r^\prime)\e{i\vec{k}\cdot\vec{r^\prime}}\phi(r^\prime) = \vspace{4pt} \\
  	=1+ \frac{1}{(2 \pi)^{2}}\int d^2r^\prime\int d^2k'\frac{\e{i(\vec{k}'-\vec{k})\cdot(\vec{r}-\vec{r}')}}{k^{\prime2}-k^2-i\epsilon}U(r^\prime)\phi(r^\prime)=1+I(r) ~.
\end{array}
\end{equation}

\noindent Choosing $\vec{k}=(k,0)$ and $\vec{k}'=(k\cos(\theta),k\sin(\theta)$), the momentum transfer vector $\vec{q}=\vec{k}'-\vec{k}$ is just $\vec{q}=(0,k\theta)$ for small angles. Changing integration variables from $k'$ to $q$ in the latter equation gives

\begin{equation}
\label{iofr}
 I(r)= \frac{1}{(2 \pi)^{2}}\int d^2r^\prime\int d^2q\frac{\e{i\vec{q}\cdot(\vec{r}-\vec{r}')}}{2\vec{k}\cdot\vec{q}+\vec{q}^{\,2}-i\epsilon}U(r^\prime)\phi(r^\prime) ~,
\end{equation}

\noindent and since $|\vec{q}| \ll 1$ we can expand $I(r)$ to a power series in terms of $|\vec{q}|^2$

\begin{equation}
 \frac{1}{2\vec{k}\cdot\vec{q}+\vec{q}^{\,2}-i\epsilon}\approx\frac{1}{2\vec{k}\cdot\vec{q}-i\epsilon}-\frac{1}{(2\vec{k}\cdot\vec{q}-i\epsilon)^2}q^2+\cdots
\end{equation}

\noindent As a result of this expansion the transmitted part of the wave function and the scattering amplitude can be written as a series as well 

\begin{equation}
\begin{array}{c}
 \phi(r)=\phi^{(1)}+\phi^{(2)}+\dots\\
 f(k,\theta)=f^{(1)}+f^{(2)}+\dots.
\end{array}
\end{equation}

\noindent Now we turn to calculate $I(r)$ to first order, bearing in mind that we chose the incident wave to propagate along the $\hat{x}$ axis, hence giving

\begin{equation}
\label{}
\begin{array}{c}
	I^{(1)}(r) = \frac{1}{(2 \pi)^{2}} \int d^2r^\prime\int dq_xdq_y\frac{\e{i(q_x(x-x')+q_y(y-y'))}}{2kq_x-i\epsilon}U(r^\prime)\phi(r^\prime) = \vspace{4pt}\\
  	 = \frac{1}{(2 \pi)}\int d^2r^\prime \int dq_x \frac{\e{iq_x(x-x')}}{2kq_x-i\epsilon}\delta(y-y')U(r^\prime)\phi(r^\prime) = \vspace{4pt}\\
	=\frac{i}{2k}\int d^2r^\prime\Theta(x-x')\delta(y-y')U(r^\prime)\phi(r^\prime) = \frac{i}{2k}\int\limits_{-\infty}^x dx^\prime U(x',y)\phi(x',y) ~.
\end{array}
\end{equation}

\noindent Using the latter result we can evaluate $\phi(r)$ and the wave function 

\begin{equation}
\begin{array}{c}
 	\phi(r)=1+\frac{i}{2k}\int\limits_{-\infty}^x dx^\prime U(x',y)\phi(x',y) = \e{\frac{i}{2k}\int\limits_{-\infty}^x dx^\prime U(x',y)} ~, \vspace{4pt}\\
 		\psi_k(r)  \approx \mbox{e}^{i(\vec{k}\cdot\vec{r}+\frac{1}{2k}\int\limits_{-\infty}^{x} dx' U(x',y))} ~.
\end{array}
\end{equation}

Since we consider here asymptotic scattering, we need to evaluate Green's function under the approximation $|r|\ll |r'|$ and hence

\begin{equation}
\label{asymp}
 |\vec{r}-\vec{r'}|=\sqrt{\vec{r}^2-2\vec{r}\cdot\vec{r}\,'+{\vec{r}\,'}^2}\approx r\sqrt{1-2\frac{\hat{r}}{r}\cdot\vec{r}\,'}\approx r-\vec{r}\, '\cdot\hat{r} ~,
\end{equation}

\noindent and then, from Eq. (\ref{psi}), the wave function can be written as

\begin{equation}
\label{ }
	\psi_{k}(r) = \mbox{e}^{i\vec{k}_{i}\cdot \vec{r}}  + \frac{\mbox{e}^{i(kr + \pi/4)}}{\sqrt{8\pi k r}}\int d^{2}r' \mbox{e}^{-i\vec{k}_{f}\cdot \vec{r}'}U(r')\psi_{k}(r') ~,
\end{equation}

\noindent where $k_i=k\hat{x}$ and $k_f=k\hat{r}$ is defined to be the scattered wave. Identifying the scattering amplitude from the asymptotic behavior of the wave function

\begin{equation}
\label{ }
	\psi_{k}(r) = \eexp{i\vec{k}\cdot\vec{r}} + \frac{1}{\sqrt{r}}f(\theta)\eexp{i(kr + \pi/4)} ~, 
\end{equation}

we get for the scattering amplitude

\begin{equation}
\label{eik1}
 f(k,\theta)=\frac{1}{\sqrt{8 \pi k}}\int d^2r'\e{i(\vec{k}_f-\vec{k}_i)\cdot\vec{r}\,'}U(r')\e{\frac{i}{2k}\int\limits_{-\infty}^{x'} dx''U(x'',y')} ~.
\end{equation}

\section{comparison of the eikonal approximation with previous results}

In this section we apply the eikonal approximation to two cases, a square well potential and a magnetic field with Gaussian ditribution, which were addressed in \cite{2d} and compare the results obtained in \cite{2d} by using Born's approximation with the new method presented here. 

\subsection{A Solenoid Magnet}

Let us consider a magnetic field generated by an ideal solenoidal current which is described by a step function realizing a uniform magnetic field pointing in the $\hat{z}$ direction and constrained to a cylindrical region around the origin

\begin{equation}
\label{ }
	\vec{B}(r) = \begin{cases} 
			B_{0} \hat{z}~,& r < R~,\cr 
			0~,& r > R~.
		       \end{cases}	
\end{equation}

\noindent Thus, the potential associated with the square well is $U(x,y)=\omega gB_{0}\Theta(x^2+y^2-R^2)$, where $B_{0}$ is the strength of the magnetic field, $g$ is the coupling constant, and $\omega$ is the energy of the incident wave. Then, we obtain

\begin{equation}
\frac{i}{2k}\int_{-\infty}^{x'} dx'' U(x'',y') = \frac{igB}{2}(x'+\sqrt{R^2-{y'}^2}) ~.
\end{equation}

\noindent Using the current limits on the axion-photon coupling constant (i.e $g \lesssim 10^{-19} ~\mbox{eV}^{-1}$ \cite{cast}), the energy of solar axions ($4.2$ eV) and current limits on terrestrial magnetic fields ($< 10^{2}$ T in the most extreme cases), the condition $|V|/E\ll 1$ is obviously satisfied. These values for the parameters will be used throughout the rest of this work.

For this potential the 2D scattering amplitude, Eq. (\ref{eik1}), is
 
\begin{equation}
 f(k,\theta)=\frac{\omega gB_{0}}{\sqrt{8\pi k}}\int dx\,'dy\,'\e{i(\vec{k}_f-\vec{k}_i)\cdot \vec{r}\,'}\e{\frac{igB_{0}}{2}(x\,'+ \sqrt{R^2 - y'^2})} ~,
\end{equation}

\noindent where the integration is preformed over the scattering region. Following the procedure from the previous section, we evaluate $\vec{k}_f-\vec{k}_i$ for small scattering angles (i.e  $\vec{q} = \vec{k}_f-\vec{k}_i \approx (0,k\theta)$)  to get

\begin{equation}
\label{fkt}
 	f(k,\theta)=\frac{\omega gB_{0}}{\sqrt{8\pi k}}\int\limits_{-R}^R dx\,'\e{\frac{i}{2}gB_{0}x\,'}\int\limits_{-\sqrt{R^2-{x'}^2}}^{\sqrt{R^2-{x'}^2}}dy\,'\e{ik\theta y' + \frac{i}{2}gB_0\sqrt{R^2-{y'}^2}} ~.
\end{equation}

\noindent In order to calculate the total cross section we shall use the optical theorem in 2D. Since the latter equation is continuous in $\theta$, one can take $\theta = 0$. We can further simplify this integral by expanding the exponential to a series in powers of $gB_0 R$. For reasonable values for a terrestrial length scale of the scattering region and the same values mentioned above for the coupling constant $g$ and magnetic field strength $B_{0}$ the first order approximations for the exponent can indeed be justified. This expansion would be done at the end of the calculation in order to have a result which is comparable to the Born approximation calculation, hence we might as well do it now. Hence, we obtain

\begin{equation}
\begin{array}{c}
 	\sigma_{tot}^{well}  = 2\sqrt{\frac{2 \pi}{k}}\mbox{Im}\{f(k,0)\} \approx \frac{1}{2}g^2B^2\int\limits_{-R}^R dx\,'\int\limits_{-\sqrt{R^2-{x'}^2}}^{\sqrt{R^2-{x'}^2}}dy\,'(x'+\sqrt{R^2-{y'}^2}) = \vspace{12pt} \\
 	= \frac{(g B_0)^2 R^3}{2} \int_{-1}^{1} d \xi\int_{-\sqrt{1-\xi^2}}^{\sqrt{1-\xi^2}} d \eta\left(\xi +\sqrt{1-\eta^2}\right) =\vspace{12pt} \\
	= \frac{8}{3} \frac{(g B_0)^2 R^3}{2} ~. 
\end{array}
\end{equation}

To obtain the conversion probability, we calculate the ratio between the number of axions arriving at the solenoid and the number of photons produced in the conversion process. The number of axions hitting the solenoid is given by multiplying the flux of incoming axions by the geometrical cross section of the solenoid, given by $\sigma_{G} = 2RL$, where $L$ is the solenoid length. In order to get the 3D total cross-section (i.e the scattering cross-section) $\sigma_{S}$ we multiply the 2D cross-section $\sigma_{tot}$ by the length of the solenoid $L$. The number of produced photons is found by multiplying the scattering cross section ($\sigma_{S} = \sigma_{tot}\cdot L$) times the flux. Thus, the conversion probability is given by 

\begin{equation}
\label{ }
	P_{well} =\sigma_{S}/\sigma_{G} = \frac{4}{3} \frac{g^{2}B_{0}^{2}R^{3}}{2R} =  \frac{2}{3} g^{2}B_{0}^{2}R^{2}~.
\end{equation} 

Comparing this result to the result obtained by using the Born approximation in \cite{2d} for the same setup (Eq. (3.20) there) we get precisely

\begin{equation}
\label{ }
	P^{eikonal}_{well}/P^{Born}_{well} = 1 ~.
\end{equation}

\noindent Hence, there is a complete correspondence between the eikonal approximation and the Born approximation in this case.

\subsection{Gaussian Magnetic Field}

In this setup the potential has the form

\begin{equation}
 U(x,y) = \omega gB_{0}\e{-\frac{x^2+y^2}{R^2}} ~.
\end{equation}

\noindent Integrating the potential along the axis of the incident wave yields

\begin{equation}
 \frac{i}{2k}\int_{-\infty}^{x'}dx''U(x'',y')  = i\frac{gB_{0}}{2}\frac{\sqrt{\pi} R}{2}[1+\mbox{Erf}(\frac{x'}{R})]\e{-\frac{y'^2}{R^2}} ,
\end{equation}

\noindent where $\mbox{Erf(x)}=\frac{2}{\sqrt{\pi}}\int_{0}^{x}\e{t^2}dt$. The scattering amplitude is given by

\begin{equation}
\begin{array}{c}
	f(k,\theta) = \sqrt{\frac{\omega}{8\pi}}gB_{0}\int_{-\infty}^{\infty} dy'\e{ik\theta y'}\e{-\frac{y'^2}{R^2}}\e{\frac{i}{4}gB_{0}R\sqrt{\pi}\e{-\frac{y'^2}{R^2}}}\int_{-\infty}^{\infty} dx'\e{-\frac{x'^2}{R^2}}\e{\frac{i}{4}gB_{0}R\sqrt{\pi}\mbox{Erf}(\frac{x'}{R})\e{-\frac{y'^2}{R^2}}} = \vspace{4pt} \\
= \sqrt{\frac{\omega}{8\pi}}4\int_{-\infty}^{\infty} dy'\e{ik\theta y'}\e{\frac{i}{4}gB_{0}R\sqrt{\pi}\e{-\frac{y'^2}{R^2}}}\sin(\frac{1}{4}gB_{0}R\sqrt{\pi}\e{-\frac{y'^2}{R^2}}) ~. 
\end{array}
\end{equation}

\noindent In order to calculate the total cross section we use, as usual, the optical theorem in 2D and by using the same reasoning as in Eq. (\ref{fkt}) to consider only the $\theta = 0$ angle we get

\begin{equation}
\label{gauss}
 \sigma_{tot}^{Gauss} = 4\int_{-\infty}^{\infty} dy'\sin^2(\frac{1}{4}gB_{0}R\sqrt{\pi}\e{-\frac{y'^2}{R^2}}) ~.
\end{equation}

\noindent To obtain an analytic result and simplify the calculation, we can use the fact that $\exp\{-\frac{y'^2}{R^2}\}\le 1$ for all $y$ and that reasonable values of the parameters $g$, $B_{0}$ and $R$ allow us to make a first order approximation. Thus,  Eq. (\ref{gauss}) can be written as

\begin{equation}
 \sigma_{tot}^{Gauss} \approx \frac{\pi}{4}g^2 B_{0}^2 R^2\int_{-\infty}^{\infty} dy'\e{-\frac{2y'^2}{R^2}}=\frac{\pi^\frac{3}{2}}{\sqrt{32}}g^2 B_{0}^2 R^3 ~.
\end{equation}

\noindent Hence, using the same method we used in the previous setup, the probability will be

\begin{equation}
 P_{Gauss}=\frac{\pi^\frac{3}{2}}{8\sqrt{2}}g^2 B_{0}^2 R^2.
\end{equation}
\noindent The comparison to the probability of conversion calculated in \cite{2d}, using the Born approximation, (Eq. (3.13) there) gives
\begin{equation}
 P_{Gauss}^{eikonal}/P_{Gauss}^{Born}=1,
\end{equation}

\noindent so that there is again a complete correspondence between the eikonal approximation and the Born approximation. Thus, we conclude that the Eikonal approximation is indeed valid and will most probably yield a correct results under our assumptions for a high energy, yet non-relativistic scattering.

\section{Axions Scattering In a Quadrupole Magnet}

After verifying the accuracy of the eikonal approximation for two known problems, we now turn to calculate the scattering of axions from a quadrupole magnetic field.

Placing the quadrupole magnet in the $yz$ plane with the $x$ axis (the direction of the incoming beam) along the symmetry axis of the quadrupole field, the quadrupole magnetic field distribution can be approximated (for a quadrupole magnet with a narrow aperture compared to its length) by \cite{stephan}

\begin{equation}
\label{ }
	\vec{B}(x,y,z) = s\vec{\nabla}({yz}) = s(z\hat{y} + y\hat{z}) = \frac{2B_{0}}{R}(z\hat{y} + y\hat{z}) ~,
\end{equation}

\noindent where $s = \tfrac{2B_{0}}{R}$ is the quadrupole gradient, $B_{0}$ is the value of the magnetic field at the pole tips (i.e at the points $(0,\pm \tfrac{R}{2},0)$ and $(0,0,\pm \tfrac{R}{2})$), where we chose a rectangular aperture for simplicity. However, since in our formalism we chose to analyze the case where the magnetic field is pointing in the $z$ direction, only the $z$ component of the external magnetic field will take part in the scattering process and we effectively have an inhomogeneous magnetic field of the form $B_{z}=\tfrac{2B_{0}}{R}y$. However, the $y$ component of the magnetic field will, of course, give the same contribution to the scattering process as the $z$ component with the sole difference that the final photons will be with a $y$-polarization. Thus, we need to take into account the $\Psi$ particle that has the $y$-polarization of the vector potential as one of its conjugate fields. This will be done at the end of this section.

Defining, in this case, the potential as:

\begin{equation}
\label{ }
	U = \omega g \frac{2B_{0}}{R}y~,
\end{equation}

\noindent for $-R/2 \leq y \leq R/2$ and $-L/2 \leq x \leq L/2$, we get

\begin{equation}
\label{ }
	\frac{i}{2k}\int_{-\infty}^{x'}dx'' U(x'',y') = ig\frac{B_{0}}{R}y'(x' + L/2)~.
\end{equation}

\noindent Putting this into Eq. (\ref{eik1}) we have for the scattering amplitude

\begin{equation}
\label{ }
\begin{array}{c}
	f(\theta)  
	= \sqrt{\frac{\omega}{2\pi}}\frac{gB_{0}}{R}\int_{-R/2}^{R/2} dy' \eexp{ik\theta y'}y' \eexp{ig\frac{B_{0}L}{2R}y'} 	\int_{-L/2}^{L/2} dx' \eexp{ig\frac{B_{0}}{R}y'x'} = \vspace{4pt} \\
	=  2 \sqrt{\frac{\omega}{2\pi}}\int_{-R/2}^{R/2} dy' \eexp{i(g\frac{B_{0}L}{2R}+ k\theta)y'} \mbox{sin}				(\tfrac{gB_{0}Ly'}{2R}) = \vspace{4pt}  \\
	= 2\sqrt{\frac{\omega}{2\pi}} \cdot i \cdot \left( \frac{\mbox{sin} (\tfrac{1}{2}kR\theta)}{k\theta} - \frac{R~\mbox{sin}(\tfrac{1}{2}(gB_{0}L + kR\theta))}{kR\theta + gB_{0}L}\right)~.
\end{array}
\end{equation}

Therefore, using the optical theorem, we obtain the total cross section 

\begin{equation}
\label{ }
	\sigma_{tot}^{quad} = \sqrt{\frac{8 \pi}{k}}\mbox{Im}\{f(0)\} = 2R \left(1- \frac{\mbox{sin}(\tfrac{1}{2}gB_{0}L)}{\tfrac{1}{2}gB_{0}L}\right) ~.
\end{equation}

To get the probability, in this case we just divide by $R$ (geometrical cross section = $R^{2}$)

\begin{equation}
\label{pquad}
	P_{quad} = 2(1- \frac{\mbox{sin}(\tfrac{1}{2}gB_{0}L)}{\tfrac{1}{2}gB_{0}L}) \approx 2\frac{\tfrac{1}{8}(gB_{0}L)^{3}}{3gB_{0}L} = \frac{1}{3}\frac{g^{2}B_{0}^{2}L^{2}}{4} ~.
\end{equation}

\noindent Although, as was discussed before, this result may be different than the one that would be obtained by using Born's approximation (since the potential is not piecewise continuous), it coincides with a result that is obtained by using an optical analogue as was previously shown by J. Redondo \cite{red}. However, Redondo's calculation is missing features of the 2D calculation (and is also computed in an unphysical setup), as we now show (and as will be discussed in the conclusions of this work).
 
The scattering from the $z$ component of the magnetic was merely a matter of choice in the in the initial setup of our system. The $\psi$ field will scatter, of course, from the $y$ component as well since this component of the magnetic is also perpendicular to the momentum of the incoming beam. The same process described above can be repeated by a $\pi/2$ rotation of the system in the $yz$ plane. In this case, the $\psi$ field would have been defined as $\tilde{\psi} = (\phi + i \tilde{A})/\sqrt{2}$, where $\tilde{A}$ is the $y$-polarization of the photon this time. This will give the same expression for the cross-section and conversion probability and since these two processes are distinguishable we can sum incoherently the two probabilities. Hence, in order to get the complete probability we have to multiply Eq. (\ref{pquad}) by a factor of two and thus

\begin{equation}
\label{39}
	P_{quad}^{total} = 2 \cdot P_{quad} =  \frac{2}{3}\frac{g^{2}B_{0}^{2}L^{2}}{4} = \frac{g^{2}s^{2}R^{2}L^{2}}{24} ~.
\end{equation}

\section{Conclusions}

In this paper we have calculated the axion-photon conversion probability from a quadrupole magnetic field. We have used the eikonal approximation to calculate the scattering from the quadrupole magnet which simplifies the calculations compared to the Born approximation and verified this approximation by comparing to known results obtained with the Born approximation. The comparison to the step function and Gaussian distributed fields shows that the eikonal approximation and the Born's approximation give reasonably close answers.

In a previous letter by J. Redondo \cite{red}, a related analysis concerning the evolution of an axion-photon field in the presence of a magnetic field with a constant gradient over the entire space, was studied. In our case, however, the constant gradient field exists only in a finite region of space. This boundary condition for the magnetic field contributes in an essential way to the scattering amplitude since, as was explained in \cite{2d}, even for the case of a strictly constant magnetic field living in a finite region of space one gets a non trivial scattering amplitude due to these boundary conditions. Thus, we conclude that the boundary effects can be as important as the gradient of magnetic field inside the finite scattering region. This is a significant difference between our treatment and that of reference \cite{red}. In Redondo's letter, the applied magnetic field does not satisfy Maxwell's equations without sources since, in his letter, $\vec\nabla \times \vec B = \partial_x B_y = B_1 = J_z$ (where in Redonodo's notations $B_y = B_1 x$). Therefore, in his analysis there is an infinite extent of currents along the $z$ axis which can simply not represent a physical situation. In conclusion, the work of Redondo is not a scattering problem and therefore cannot incorporate all the physical aspects of an experimental set-up as opposed to the work presented here. This is most easily seen by comparing Eq. (\ref{39}) with equation 3 in Redondo's letter. This comparison unveils the fact that the different approaches lead to different results.

Moreover, it may sometimes be tempting to believe that the 2D results can be obtained by averaging over 1D conversion probabilities. As was already shown in \cite {2d}, a general prescription to find a 1D analogue to the 2D calculation may be obtained by using the magnetic flux as the averaging measure

\begin{equation}
\label{avg}
	P_{1D}^{avg} = \frac{\int_{-\infty}^{\infty}\int_{-\infty}^{\infty}\tfrac{1}{4}g^2 |\int_{-\infty}^{\infty}B(x',y)dx'|^2 B(x,y)dxdy}{\int_{-\infty}^{\infty}\int_{-\infty}^{\infty}B(x,y)dxdy} ~.
\end{equation}

\noindent However, since the quadrupole magnetic flux is zero, this method will not work this time. Of course, it is possible to find an averaging process that will produce the 2D result, but it cannot be done a-priori with certainty. Eq. (\ref{avg}) is an example for a legitimate choice of measure that cannot produce a result at all. It is clear that there is no way to avoid the real calculation and obtain results in higher dimensions from averaging on lower dimension estimations. This shows that 2D processes cannot be reduced to a 1D calculation. The scattering process from a quadrupole field is intrinsically 2D since the scattered photon may have two different polarizations: Considering a magnetic field produced by a \textit{physical} localized current naturally makes a significant difference. In particular, it shows that any attempt to say that the problem can be deduced from a 1D analogue is untenable and, in fact, will fail since the magnetic field produced by a physical current will necessarily have at least two components (in the source free region). Hence, in the process of photon production from the scattering of axions, the photon will have two distinguished polarizations. In addition, in the source free regions, a non-uniform field pointing in one direction, of the form $\vec B = B(x,y)\hat l$, will not be able to satisfy the source free Maxwell's equations, $\vec\nabla \times \vec B = \vec\nabla B(x,y) \times \hat l =0$ and $\vec\nabla \cdot \vec B = \vec\nabla B(x,y)\cdot \hat l = 0$, since the solution requires that $\vec\nabla B(x,y) = 0$. Therefore, a one directional magnetic field cannot solve these equations and the problem cannot be reduced to 1D.

We can, however, compare the scattering from a quadrupole to the scattering from a solenoid, which can represent a dipole accelerator magnet (with a different geometry of course). The result obtained here shows that it will be preferable to have a constant magnetic field distributed over the scattering region aperture (like, for example, the field of a solenoid) rather than having an inhomogenous field (as the quadrupole field). This comes from the fact that the magnetic field energy will be higher in the first case (when the maximal magnetic field strength $B_{0}$ is equal for both cases and both fields are distributed over the same scattering region). Since $B_{0}$ is an intrinsic property of the superconducting material, comparing a quadrupole magnet to a dipole magnet of the same length and aperture and made with the same superconductor, the dipole will yield a higher conversion rate.

One can also observe that the magnetic field parameter that determines the conversion rate is actually the global maximum value of the magnetic field in the scattering region. This feature appears as well in other calculations of the conversion probability from inhomogeneous fields, like, for example, the Gaussian distributed field in \cite{2d}.

\vskip.6in

\centerline{{\bf Acknowledgments}} We wish to thank K. Zioutas for helpful discussions and comments. I.S would like to thank the CAST collaboration for their kind hospitality and support and to the MSC/TE group for their attentive hospitality during his stay at CERN.

\vskip.3in


\end{document}